\documentstyle[preprint,revtex]{aps}
\begin{document}
\draft
\baselineskip = 1.0\baselineskip
\begin{title}
{On free energy of three--dimensional Ising model at criticality}
\end{title}
\author{ A. I. Sokolov,~~V. A. Ul'kov,~~E. V. Orlov }
\begin{instit}
Department of Physical Electronics, Saint Petersburg 
Electrotechnical University, \\
Professor Popov Street  5, St.Petersburg 197376, Russia
\end{instit}

\begin{abstract}

Higher--order vertices at zero external 
momenta for the scalar field theory, describing the critical 
behaviour of the Ising model, are studied within the 
field--theoretical renormalization group (RG) approach
in three dimensions. Dimensionless six--point $g_6$ and 
eight--point $g_8$ effective coupling constants are calculated 
in the three--loop approximation. 
Their numerical values, universal at criticality, are estimated 
by means of the Pade and Pade--Borel summation of the RG 
expansions found and by putting the renormalized quartic 
coupling constant equal to its universal fixed--point value 
known from six--loop RG calculations. The values 
of $g_6^*$ obtained are compared with their analogs resulting 
from the $\epsilon$--expansion, Monte Carlo simulations, the 
Wegner--Houghton equations and the linked cluster expansion 
series. The field--theoretical estimates for $g_6^*$ are shown 
to be in a good agreement with each other, 
differing considerably from the values given by other methods.

\end{abstract}

\newpage

\section{Introduction}

The critical thermodynamics of the three--dimensional Ising 
model is known to be described by Euclidean scalar field theory 
with the Hamiltonian
\begin{eqnarray}
H = 
\int d^3x \Bigl[{1 \over 2}( m_0^2 \varphi^2
 + (\nabla \varphi)^2) 
+ {\lambda \over 4!} \varphi^4 \Bigr] ,
\label{eq:1} 
\end{eqnarray} 
where a bare mass squared $m_0^2$ is proportional to 
$T - T_c^{(0)}$, $T_c^{(0)}$ being the phase transition 
temperature in the absence of the order parameter fluctuations. 
Taking fluctuations into account leads to renormalization of 
the mass $m_0^2 \to  m^2 $, the field $\phi \to  \phi_R $, and 
the coupling constant $\lambda \to  mg_4$, and also to the 
appearance of terms of the form $m^{3-n} g_{2n} \phi_R^{2n}$ 
with $n > 2$ in the effective action (free energy) of the 
system. In the critical region, where fluctuations are 
so strong that they completely screen out the initial (bare) 
interaction, the behaviour of the system becomes universal and 
the dimensionless vertices $g_{2n}$ tend toward their 
asymptotic limits, i.e. they assume constant 
values which are also universal.

In this paper, we calculate numerical values of $g_6$ and $g_8$ 
at criticality using the field--theoretical renormalization group 
(RG) approach in three dimensions. Higher--order coupling 
constants mentioned related to corresponding vertices at zero 
external momenta will be found as power series in the 
renormalized dimensionless quartic coupling constant $g_4$ up to 
three--loop order, and the RG series will be resummed by means 
of the Pade and Pade--Borel techniques. Then $g_4$ in resummed 
RG expansions will be put equal to its universal 
value $g_4^*$ known from the canonical six--loop RG calculations 
resulting in numerical estimates for $g_6^*$ and $g_8^*$.
The numbers obtained for $g_6^*$ will be compared with those 
resulting from the $\epsilon$--expansion for $g_6 / g_4^2$ as 
well as with the values given by Monte Carlo simulations, by the 
approximate solution of the Wegner--Houghton equations and by the 
analysis of the linked cluster expansions.  

\section{RG expansions for sextic and octic effective interactions}

Since higher--order bare coupling constants are known to be 
irrelevant at criticality in the RG sense, renormalized 
perturbative expansions for $g_6$, $g_8$, etc. may be obtained 
from conventional Feynman--graph expansions of these quantities 
in terms of the only bare coupling constant -- $\lambda$. 
In its turn, $\lambda$ may be expressed perturbatively as a 
function of the renormalized dimensionless quartic coupling 
constant $g_4$. Substituting corresponding power series for 
$\lambda$ into original Feynman--graph expansions for $g_6$ 
and $g_8$, we can obtain the RG series for 
these higher--order effective coupling constants.   

Recently, one of the authors has found the sextic coupling 
constant $g_6$ in the two--loop RG approximation \cite{ais}. 
Thus, what we are interesting in is the three--loop 
contribution to this quantity. As may be shown, there are 16 
graphs contributing to $g_6$ in the three--loop order. In fact, 
their calculation is neither cumbersome nor lengthy since 
corresponding contribution may be written down as a sum of 
a few terms having a form of mass derivatives of some two--point 
and four--point graphs. So, within the three--loop 
approximation we get

\begin{eqnarray}
g_6 = {9 \over \pi}{{\Bigl({g_0 Z^2 \over m} \Bigr)}^3}{\Bigl
[ 1 - {33 \over 2 \pi}{g_0 Z^2 \over m} + 
20.53966666 {\Bigl({g_0 Z^2 \over m} \Bigr)^2} \Bigr]} ,
\label{eq:2} 
\end{eqnarray} 
where $g_0 = \lambda / 4!$ and $Z$ relates the dressed Green 
function $G$ to the renormalized one $G_R$ in a conventional way:
\begin{eqnarray}
G = Z{G_R} = {Z \over {m^2 + q^2 - \Sigma_R}} = {Z \over {m^2 + 
q^2}} + O(g_4^2).
\label{eq:3} 
\end{eqnarray}

The renormalized perturbative expansion for bare coupling 
constant $g_0$ may be obtained using the normalizing condition 
\begin{eqnarray}
g_0 = {m g_4}{Z_4 \over {Z^2}}
\label{eq:4}
\end{eqnarray}
and the RG expansion for $Z_4$ which has been calculated by 
many people:
\begin{eqnarray}
Z_4 = 1 + {9 \over {2 \pi}}g_4 + {63 \over {4 \pi^2}}g_4^2 + 
O(g_4^3).
\label{eq:5}
\end{eqnarray} 
Combining Eqs. 2, 4 and 5 we obtain
\begin{eqnarray}
g_6 = {9 \over \pi}{g_4^3}{\Bigl( 1 - {3 \over \pi}{g_4} + 
1.389962952 {g_4^2} \Bigr)}.
\label{eq:6}
\end{eqnarray}

The RG expansion for the eight--point effective coupling 
constant $g_8$ may be found in an analogous way although it 
requires more job than in the case of $g_6$. 
Two--loop and three--loop contributions to $g_8$ are given 
by 5 and 36 Feynman graphs respectively. Use of the trick 
mentioned above, however, considerably simplifies their 
calculation. The result is as follows 
\begin{eqnarray}
g_8 = - {81 \over 2 \pi}{{\Bigl({g_0 Z^2 \over m} \Bigr)}^4}
{\Bigl[ 1 - {173 \over 6 \pi}{g_0 Z^2 \over m} +
54.81336082 {\Bigl({g_0 Z^2 \over m} \Bigr)^2} \Bigr]}.
\label{eq:7}
\end{eqnarray}
Expressing $g_0$ in terms of $g_4$ leads to the RG series 
for $g_8$:
\begin{eqnarray}
g_8 = - {81 \over 2 \pi}{g_4^4}{\Bigl( 1 - {65 \over 6 \pi}{g_4} 
+ 7.775001310 {g_4^2} \Bigr)}.
\label{eq:8}
\end{eqnarray}   

With the RG expansions Eqs. 6, 8 in hand, we can get numerical 
estimates for universal critical values of $g_6$ and $g_8$. 

\section{Resummation and numerical estimates}

Perturbative expansions in a field theory are known to be 
divergent, at best asymptotic. Moreover, the theory under 
consideration has no small parameter. That is why direct 
substitution of the fixed point value $g_4^*$ to Eqs. 6, 8 
can not lead to satisfactory results.
These expansions, however, contain important information which 
may be extracted provided some procedure making them convergent 
is applied. Here Pade and Pade--Borel methods will play roles 
of such procedures, i.e. we shall construct Pade approximants 
$[L/M]$ for the functions given by the series Eqs. 6, 8 as well 
as for their Borel transforms which are related to the functions 
to be found ("sum of series") by the formula
\begin{eqnarray}
f(x) = \sum_{k = 0}^{\infty} c_k x^k = \int\limits_0^{\infty}
e^{-t} F(xt) dt \ \ ,
\label{eq:9}
\end{eqnarray}
\begin{eqnarray}
F(y) = \sum_{k = 0}^{\infty} {c_k \over {k!}} y^k \ \ ,
\label{eq:10}
\end{eqnarray}
and then evaluate the integral Eq. 9 where series Eq. 10 
possessing nonzero radii of convergence are replaced by 
corresponding Pade approximants.

Starting from the three--loop expansions available, it is 
possible to construct Pade approximants of the only reasonable 
type: $[1/1]$. On the other hand, one can try several different 
ways of resummation. Indeed, the Pade--Borel procedure may be 
applied not only to the series for $g_6$ and $g_8$ themselves 
but also to corresponding RG expansions 
for the ratios $g_6 / g_4^2$ and $g_8 / g_4^3$.

Let's construct first the Pade approximant for the six--point 
effective coupling constant. From Eq. 6 we readily obtain:
\begin{equation}
g_6 = {9 \over \pi}{g_4^3}{{1 + 0.500636 g_4} \over {1 + 
1.455566 g_4}}.
\label{eq:11}
\end{equation}
Substituting into this expression the most accurate numerical 
estimate $g_4^* = 0.988$ for the fixed point value known from 
six--loop RG calculations in three dimensions \cite{{lgz},{bnm}} 
we find for the universal value of $g_6$:
\begin{equation}
g_6^* = 1.694.
\label{eq:12}
\end{equation}

Apply then more sophisticated, Pade--Borel procedure to the 
series Eq. 6 which is expected to lead to more accurate 
estimate for $g_6^*$. After simple algebra we obtain
\begin{eqnarray}
g_6 = 0.477465 {g_4^3} \int\limits_0^{\infty} 
{{1 + 0.052382 g_4 t} \over {1 + 0.291114 g_4 t}}{t^3} e^{-t} dt.
\label{eq:13}
\end{eqnarray}  
Computation of this integral for $g_4 = 0.988$ gives
\begin{equation}
g_6^* = 1.621.
\label{eq:14}
\end{equation}
The third way to get numerical estimate for $g_6^*$ we use here 
is to construct the Pade--Borel approximant for $ g_6 / g_4^2 $ 
and to estimate the universal value of this ratio at criticality.  
The corresponding expression is as follows:
\begin{eqnarray}
{g_6 \over {g_4^2}} = 2.86479 {g_4} \int\limits_0^{\infty}
{{1 + 0.0077234 g_4 t} \over {1 + 0.48519 g_4 t}} t e^{-t} dt.
\label{eq:15}
\end{eqnarray}
For the fixed point value of $g_4$ this formula leads to:
\begin{equation}
g_6^* = 1.577.
\label{eq:16}
\end{equation} 

To obtain numerical estimates for $g_8^*$ the same procedures 
are applied to the series Eq. 8. They give
\begin{eqnarray}
g_8^* &=& 0.68 \quad {\rm (Pade)}, \nonumber \\
g_8^* &=& 1.71 \quad {\rm (Pade-Borel)},  \\
g_8^* &=& 2.71 \quad {\rm (Pade-Borel \  for} \  g_8 / g_4^3). 
\nonumber
\label{eq:17}
\end{eqnarray}

\section{Discussion}

Although the RG expansions found in Sec.2 are, in fact, rather 
short, numerical estimates for $g_6$ obtained from Eq. 6 by 
several resummation techniques are seen to be close to each 
other. That is why we believe that the numbers (12) and, 
in particular, (14) and (16) are also close   
to the exact value of $g_6^*$. Moreover, taking into account 
of the three--loop RG contribution to $g_6^*$ turns out to 
change corresponding estimate by less than 10 per cent 
(Pade--Borel resummation of the two--loop RG expansion have 
lead to $g_6^* = 1.50$ \cite{ais}) what may be considered as 
an extra argument in favour of fair numerical accuracy 
of the results presented above.

Let's compare our estimates for $g_6^*$ with their counterparts 
obtained by different methods. The solution of the 
Wegner--Houghton equations within the local potential 
approximation presented by C.Bagnuls and C.Bervillier
has yielded $(g_6 / g_4^2) = 3.59$, $g_6 = 2.40$ at the 
non--trivial fixed point \cite{bb}. Determining by Monte Carlo 
simulations probability distributions for averaged magnetization 
of the 3D Ising model in an external magnetic field, M.M. Tsypin 
has found from his data that $g_6^* = 2.05$ \cite{mmt}. 
The analysis of the linked cluster expansion series performed 
by T.Reisz has lead to $g_6^* = 1.92$ \cite{tr}. 

All these estimates are seen to be significantly larger than 
ours obtained by means of perturbative RG calculations for 
the field--theoretical model. In such a situation, it is 
interesting to evaluate $g_6^*$ using some alternative 
field--theoretical perturbative approach. 
The $\epsilon$--expansion for the ratio $g_6 / g_4^2$ which 
results from the equation of state may be 
used for this purpose. Up to the ${\epsilon}^3$ order, it is as 
follows \cite{jz}:
\begin{equation}
{g_6 \over {g_4^2}} = 2 \epsilon - {20 \over {27}} {\epsilon}^2 + 
1.2759 {\epsilon}^3. 
\label{eq:18}
\end{equation}   
Resumming this expansion by Pade and Pade--Borel methods we find
for $\epsilon = 1$ and $g_4 = 0.988$:
\begin{eqnarray}
g_6^* &=& 1.687 \quad {\rm (Pade)}, \nonumber \\
g_6^* &=& 1.653 \quad {\rm (Pade-Borel)}.
\label{eq:19}
\end{eqnarray}

The Pade--Borel estimate for $g_6^*$ thus obtained is remarkably 
close to those given by RG calculations in three dimensions 
(Eqs. 14 and 16). So, the field--theoretical RG approach in three 
and $(4 - \epsilon)$ dimensions provides, within the three--loop 
approximation, the values of $g_6^*$ which agree well to each 
other but differ considerably from the Monte Carlo and some other 
estimates mentioned above.

Let's discuss further the predictions concerning the eight--point 
effective coupling constant. The estimates following from the 
three--loop RG expansions in 3D are cleary seen to be strongly 
scattered being therefore numerically unreliable. It is not 
surprising since coefficients in the RG expansion for $g_8$ 
grow much faster than those in the RG series for $g_6$ making 
the former series less suitable for resummation than the latter. 
Correspondingly, much less reliable numerical estimates for 
$g_8$ at the critical point are obtained. On the other hand, 
all three numbers Eqs. 17 are i) positive, i.e. have a sign 
opposite to that of the lowest--order (one--loop) RG estimate 
$g_8^* = - 12.28$, and ii) much smaller than the absolute value 
of this one--loop estimate. It means that higher--order RG 
contributions to $g_8$ being extremely important tend to strongly 
diminish its universal value at the critical point with respect 
to the number given by the lowest--order RG approximation. 
This conclusion seems to be in accord with the fact that the 
recent analysis of relevant Monte Carlo data failed to reveal 
some appreciable (non--zero) value of $g_8^*$ \cite{mmt}.  

\section{Conclusion}

In the paper, the RG series for coefficients before $M^6$ and 
$M^8$ in the expansion of the free energy of the 
three--dimensional Ising model in powers of the order parameter 
$M$ have been calculated in the three--loop approximation. 
Numerical estimates for universal values of $g_6$ 
and $g_8$ at the critical point have been found by Pade and 
Pade--Borel resummations of the series obtained and by putting 
the quartic coupling constant equal to its fixed point value 
0.988. Pade--Borel procedure applied to the expansions for $g_6$ 
and $g_6 / g_4^2$ has given $g_6^* = 1. 621$ and $g_6^* = 1.577$, 
respectively, while analogous treatment of the 
$\epsilon$--expansion for $g_6 / g_4^2$ has been shown to result 
in $g_6^* = 1.653$. Being in a good agreement to each other, 
these field--theoretical estimates are considerably smaller 
than those obtained by other methods. The RG expansion for
$g_8$ has turned out to be less suitable for resummation than 
that for $g_6$ since it possesses rapidly growing coefficients. 
As a result, strongly scattered numerical estimates for $g_8^*$
lying between 0.68 and 2.71 have been found.

\section{Acknowledgment}

It was very interesting to compare our results with those
presented by Prof. J. Zinn--Justin at the International 
Conference "Renormalization Group '96" (Dubna, Russia, 
August 1996). One of the authors (A.I.S.) cordially thanks 
him for such an opportunity and for discussion.

\end{document}